\documentclass[smallabstract,smallcaptions]{dccpaper}

\usepackage{epsfig}
\usepackage{cite}
\usepackage{amsmath}
\usepackage{amssymb}
\usepackage{color}
\usepackage{url}

\usepackage{comment,booktabs}
\usepackage[pagebackref=true,breaklinks=true,colorlinks,bookmarks=false]{hyperref}

\usepackage[caption=false]{subfig}

\newlength{\figurewidth}
\newlength{\smallfigurewidth}

\begin{document}

\title{\large
	\textbf{End-to-end optimized image compression for machines, a study}
}

\author{%
	Lahiru D. Chamain$^{\ast\dag\ddagger}$, Fabien Racapé$^{\ast\ddagger}$, Jean Bégaint$^{\ast}$, Akshay Pushparaja$^{\ast}$,\\
	and Simon Feltman$^{\ast}$\\[0.5em]
	\begin{minipage}{0.95\textwidth}
	    \small
        \centering
		\begin{tabular}{cc}
			$^{\ast}$InterDigital \-- AI Lab    & $^{\dag}$University of California, Davis \\
			4410 El Camino Real                 & 1 Shields Ave \\
			Los Altos, CA, 94022, USA           & Davis, CA, 95616, USA \\
			\{fabien.racape,jean.begaint,       & hdchamain@ucdavis.edu \\
		    akshay.pushparaja,simon.feltman\}   & \\
		    @interdigital.com   & \\ 
		\end{tabular}
	\end{minipage}
	\thanks{$\ddagger$ Equal contribution.}
	\thanks{This work was done while Lahiru D. Chamain was an intern at the InterDigital AI Lab.}
}

\maketitle
\thispagestyle{empty}

\begin{abstract}

An increasing share of image and video content is analyzed by machines rather than viewed by humans, and therefore it becomes relevant to optimize codecs for such applications where the analysis is performed remotely. Unfortunately, conventional coding tools are challenging to specialize for machine tasks as they were originally designed for human perception. However, neural network based
codecs can be jointly trained end-to-end with any convolutional neural network (CNN)-based task model. 
In this paper, we propose to study an end-to-end framework enabling efficient image compression for remote machine task analysis, using a chain composed of a compression module and a task algorithm that can be optimized end-to-end. We show that it is possible to significantly improve the task accuracy when fine-tuning jointly the codec and the task networks, especially at low bit-rates. Depending on training or deployment constraints, selective fine-tuning can be applied only on the encoder,
decoder or task network and still achieve rate-accuracy improvements over an off-the-shelf codec and task network. Our results also demonstrate the flexibility of end-to-end pipelines for practical applications.

\end{abstract}

\section{Introduction}

\begin{figure}[t]
	\centering
	\includegraphics[width=\textwidth]{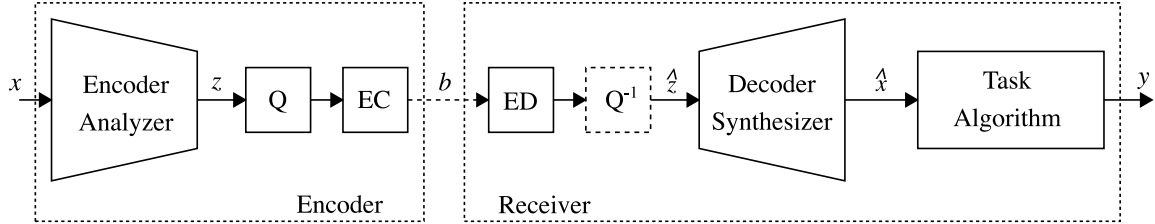}
	\caption{General structure of the proposed study. Input images $\mathbf{x}$
		are transformed into latent $\mathbf{z}$, then quantized (Q) and entropy
		coded (EC). The binary stream $b$ is then stored or transmitted over a
		channel. On the receiver side, the decoder reconstructs the input images
		as $\mathbf{\hat x}$, from the decoded latent tensors $\mathbf{\hat z}$, on
	which learning/inference can be performed by the task algorithm.}\label{fig:main}
\end{figure}

An increasing share of captured images and videos are now exclusively
	analyzed by machines, rather than viewed by humans, to perform computer vision analysis tasks such as object detection, image
	classification, segmentation, etc. These machine tasks usually require large
	CNNs which are computationally challenging and might need
	to be executed remotely. For example, embedded low-end devices can
	capture and transmit encoded videos to computing data-centers for analysis and storage. The machine tasks would then be performed on reconstructed images or videos,
	which are partially distorted by the encoding process.
	
	So far, compression standards, such as JPEG~\cite{jpeg}, H.265/HEVC~\cite{hevc},
	AV1~\cite{av1} and the recent H.266/VVC~\cite{benjamin_bross_jvet}, have
	been designed to minimize the distortion of decoded content in terms of visual quality. For
	instance, the human visual system (HVS) is less sensitive to distortions in
	the high spatial frequencies than in the lower frequencies. Traditional codecs
	transform the signal using variants of discrete sine or cosine transforms to
	to enable the encoder to quantize high spatial frequencies more severely. The HVS is
	also more sensitive to distortions on the luma component (intensity) than on
	the chroma components (colors), which motivated coding tools to favor higher fidelity
	for the luma component. However, machine task algorithms may not be accurate
	on the content that was compressed based on HVS properties. For instance,
	high spatial frequency features can be useful for object detection and
chroma components may have a significant impact when classifying the elements of
a scene. The field of Video Coding for Machines (VCM) thus has gained
	traction in recent years~\cite{duan_video_2020}.
	
Contrary to traditional hybrid video compression frameworks, artificial neural network (ANN)-based codecs can be directly optimized with respect to any rate-distortion
criterion. The distortion criterion is a differentiable loss function and is
not limited to a signal fidelity metric such as the PSNR or SSIM\@. For known
machine task architectures, prediction errors on decoded images can be used to
directly optimize the parameters of an auto-encoder. A natural extension is then
to jointly optimize an end-to-end system consisting of a trained codec and a
CNN-based task algorithm.

This paper provides an in-depth study of a framework consisting of a trained
auto-encoder and a machine task algorithm applied to the decoded
frames. An illustration of the complete pipeline is provided in
Figure~\ref{fig:main}. The output of the decoder still consists of
images that can be fed to the task algorithms, which means that existing
machine task architectures can be directly used on the decoded images. 
Although only still picture coding is considered in this paper, the studied
properties of coupled compression and machine tasks are expected to hold for
videos as well.
In this setting, the jointly optimized codec produces
decoded images that are `specialized' for the task considered during training. We show that it is possible to significantly improve the task accuracy when re-training jointly the codec and the task networks.

The remainder of the paper is organized as follows. Related work is first
discussed in Section~\ref{sec:related}. In Section~\ref{sec:study}, we detail
the different configurations of joint optimization that are explored. In
Section~\ref{sec:results}, we then provide extensive experimental
results in the context of object detection, including different
fine-tuning configurations for the different parts of the chain. Finally, we
discuss and conclude in Section~\ref{sec:conclusion} on the promises and
limitations of the studied approaches.
\section{Related works}\label{sec:related}

In recent years, multiple ANN-based codecs have been proposed for image
	compression~\cite{balle2018variational,minnen2018joint,cheng_learned_2020,minnen2020channel}, replacing conventional coding tools with end-to-end neural networks.
	Trained codecs have now caught up with conventional stat-of-the-art codecs
	such as H.265/HEVC~\cite{hevc} and AV1~\cite{av1}, in terms of
	rate-distortion performances for still picture coding configuration. We
	chose to base our study on one of the most representative approaches
	introduced by Ball\'{e} \textit{et al.} in~\cite{balle2018variational}. They proposed a method
	composed of an auto-encoder transforming the input image into a quantized
	tensor of feature maps (or latent), which is then entropy coded. A second auto-encoder, the \textit{hyperprior}, improves the compression efficiency by encoding side-information about the latent element probability distributions. Most of the recent state-of-the-art ANN-based methods now derive from this original hyperprior architecture~\cite{cheng_learned_2020,minnen2020channel,mentzer_high-fidelity_2020}.

In terms of optimizing image/video compression codecs for a given learning task, only a handful of works emerges from the literature. The authors of~\cite{luo2020rate} and~\cite{chamain2018QuanNet} used an approximate differentiable codec implementation to learn better quantization tables for JPEG and JPEG2000. The method presented in~\cite{luo2020rate} shows performance gain in rate-distortion-accuracy for ImageNet-1K by training the differentiable codec with a fixed MobileNet architecture~\cite{howard2019searching}. However, the rate-accuracy performance in~\cite{luo2020rate,chamain2018QuanNet} is still limited by coding tools designed for visual quality. While end-to-end learning of compressible features has already been proven to be efficient when targeting machine vision tasks~\cite{singh2020end}, no study has been performed on the accuracy of jointly fine-tuned CNNs and ANN-based image codecs.

\section{End-to-end optimized compression for machines}\label{sec:study}\label{sec:framework}

We propose to study the structure shown in Figure~\ref{fig:main}, which consists
of an image encoder and a receiver composed of an image decoder and a machine task algorithm.
The auto-encoder, or codec, consists of a learned encoder analyzer, a quantization
module, an entropy coder (EC), an entropy decoder (ED), an optional inverse
quantization module, and a learned decoder synthesizer. The encoder analyzer transforms
the input images $\mathbf{x}$ into a compressible, lower dimensional latent
representation $\mathbf{z}$, which is then quantized and entropy coded to form a
bit-stream $b$. The entropy decoder decodes the bit-stream to reproduce
$\mathbf{\hat z}$ which is then fed to the decoder synthesis stage to
generate the reconstructed images $\mathbf{\hat x}$. During training, to keep the chain
differentiable, the entropy coder actually estimates the probabilities of the
elements in the latent $\mathbf{z}$ to approximate the real bit-rate of each input
image~\cite{balle2018variational}.

The ANN-based method from~\cite{balle2018variational} used in this paper
requires to train a model for each rate point, or quality level $q$, of the
reconstructed images ($\mathbf{\hat x}_q$). Hence, we parameterize the codec
that encodes images at quality $q$ with $\boldsymbol \psi_q$, ($\mathbf{\hat
x}_q = \boldsymbol \psi_q(\mathbf{x})$). The `task algorithm', parameterized
with $\boldsymbol \theta$, then takes reconstructed images $\mathbf{\hat x}_q$
and estimates the task prediction $y$, ($\boldsymbol \theta(\mathbf{\hat
x}_q)=y$). The inference accuracy can be calculated by considering the
difference between the estimated label $y$ and the ground truth label $y_{gt}$
for $\mathbf{x}$. In general, the task model $\boldsymbol \theta$ is optimized
to reach a minimum task loss on the input images that are used for training,
which are usually compressed images at high quality $rq$. We denote the task
model optimized for the best quality $rq$ by $\boldsymbol \theta_{rq}$.

We explore 4 main settings regarding the possible re-training of the codec
and/or the task algorithm. The first setting consists in exploring the
rate-accuracy performance by performing the inference on an
off-the-shelf (baseline) model $\boldsymbol \theta_{rq}$ with the inputs generated by
a rate-distortion optimized codec $\boldsymbol \psi_{rq}$, \textit{i.e.} the
pre-trained models, which we will consider as baseline. In the second setting we
fine-tune the task model (T-FT), \textit{i.e.} we re-train starting from
pre-trained parameters, for each rate-distortion optimized codec at quality $q$.
Third, we explore the rate-accuracy performance by fine-tuning the codec (C-FT)
for an off-the-shelf task model. As final setting, we jointly re-train,
end-to-end, the codec and the task model (J-FT) to optimize rate-accuracy
performance of the considered task $T$.

\subsection{Inference on off-the-shelf task model: baseline}\label{sec:baseline}

Almost all the existing pre-trained deep learning models that are designed and optimized to perform
computer vision tasks such as object detection, segmentation, or tracking have
been trained on (high quality) compressed images or video frames, sometimes on raw images or videos.
In this study, the quality of the source dataset will be denoted $rq$ as it represents our reference quality level. By compressing images
($\mathbf{\hat x}_q$) at different qualities $q$, we can measure the performance
in terms of rate vs.\ task accuracy (rate-accuracy). This forms the baseline for
the studied methods, referred to as \textit{baseline} in the
remainder.


\subsection{Fine tuning the task model only: T-FT}\label{sec:ftt}

In this section, the task model is re-trained to take into account the distortions introduced by the compression.
In the following, $\mathcal{X}_{rq}$ denotes the sets containing reference quality images
$\mathbf{x}\in\mathcal{X}_{rq}$ and $\mathcal{\hat X}_q$ denotes the set of
decompressed images at quality $q$ \textit{s.t.}
$\mathbf{\hat x}_q\in\mathcal{\hat X}_q$. $y_{gt} \in \mathcal{Y}$ represent the
ground-truth predictions for the inputs $\mathbf{x}$ and $\mathbf{\hat x}_q$ for the
considered task.
We express the loss function for a task $T$ with $\mathcal{L}_T(y,y_{gt})$,
where the prediction $y=\boldsymbol \theta(\mathbf{\hat x}_q)$ is generated by
the task model and compared with the ground truth $y_{gt}$.

The original task model ($\boldsymbol\theta_{rq}$) can then be re-trained using the images
$\mathbf{\hat{x}}_q$ compressed at each quality level ($q$), to obtain a `fine tuned'
task model ($\boldsymbol\theta_q$).

\begin{align}
	\label{eq:tft}
	\boldsymbol \theta_q = \arg \min_{\boldsymbol \theta|\boldsymbol \theta_{rq}} \dfrac{1}{|\mathcal{\hat X}_q|}\sum_{\mathbf{\hat x}_q ,y_{gt} \in\mathcal{\hat X}_q, \mathcal{Y}} \mathcal{L}_T(\boldsymbol \theta(\mathbf{\hat{x}}_q),y_{gt})
\end{align}

Note that the re-training to optimize $\boldsymbol \theta$ is initialized with
the off-the-shelf $\boldsymbol \theta_{rq}$ denoted by $\boldsymbol
\theta|\boldsymbol \theta_{rq}$.

\subsection{Fine tuning the codec only: C-FT}\label{sec:cft}

Under the assumption that the graph of the task algorithm is known and the task loss
can be back-propagated through the task model, we can fine-tune the codec
$\boldsymbol \psi$ by optimizing the trained parameters of the auto-encoder,
without updating the task model $\boldsymbol \theta$. To that end, we minimize the
following loss to obtain a codec $\boldsymbol \psi_{\beta}^T$ s.t.

\begin{align}
	\label{eq:cft}
	\boldsymbol \psi_{\beta}^T = \arg \min_{\boldsymbol \psi|\boldsymbol \psi_q} \dfrac{1}{|\mathcal{X}_{rq}|} \sum_{\mathbf{x} ,y_{gt} \in\mathcal{ X}_{rq}, \mathcal{Y}} & \{\mathcal{L}_T(\boldsymbol \theta_{rq}(\boldsymbol \psi(\mathbf{x})),y_{gt}) + \beta \mathcal{L}_R(\mathbf{z})\}
\end{align}

where $\mathcal{L}_R(\mathbf{z})$ denotes the part of the loss relating to the
estimated size, or Rate (R) of the bit-stream (b), which is the output of the
encoder $\boldsymbol{\psi}$ s.t. $\mathbf{z} = \rm
E_{\boldsymbol{\psi}}(\mathbf{x})$ as shown in Figure~\ref{fig:main}. The
control parameter $\beta$ regulates the trade-off between the accuracy of task
$T$ and the rate R.

\subsection{Joint end-to-end fine tuning: J-FT}

This section describes the joint optimization of the parameters of both the
encoder $\boldsymbol \psi$ and the task model $\boldsymbol \theta$. The
optimal `codec-task model' combination ($\boldsymbol \psi_{\beta}^J,\boldsymbol
\theta_{\beta}^J$) can be learned by minimizing

\begin{align}
	\label{eq:jft}
	\boldsymbol \psi_{\beta}^J, \boldsymbol \theta_{\beta}^J = \arg \min_{\boldsymbol \psi|\boldsymbol \psi_q,\boldsymbol \theta|\boldsymbol \theta_{rq}} \dfrac{1}{|\mathcal{ X}_{rq}|} \sum_{\mathbf{x} ,y_{gt} \in\mathcal{ X}_{rq}, \mathcal{Y}} & \{\mathcal{L}_T(\boldsymbol \theta(\boldsymbol \psi(\mathbf{x})),y_{gt}) + \beta \mathcal{L}_R(\mathbf{z})\}.
\end{align}

In this scenario, we initialize the codec and the task model with pre-trained models
$\boldsymbol \psi_q$ and $\boldsymbol \theta_{rq}$ for each $\beta$ value.

\section{Experiments and Results}\label{sec:results}

\begin{figure}[t]
    \centering
	\includegraphics[trim=10 10 20 40,clip,width=0.7\linewidth]{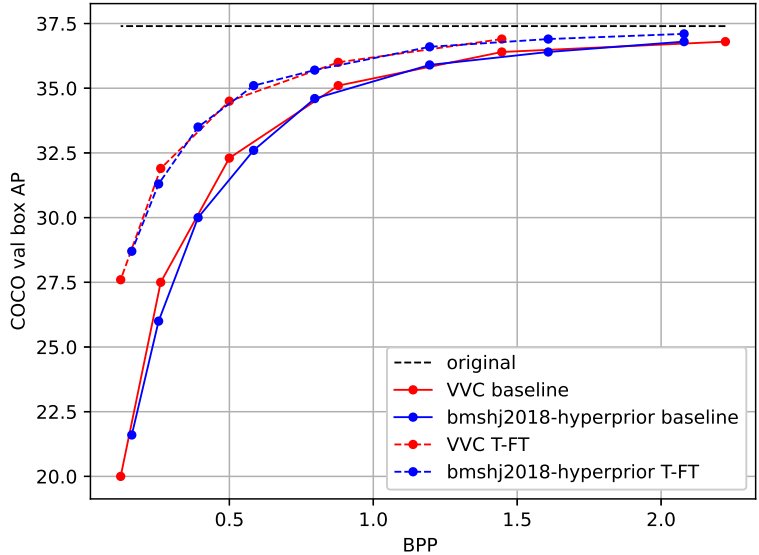}
	\caption{Rate-accuracy comparison for VVC and the \textit{bmshj2018-hyperprior} model. Solid lines represent the baseline, whereas dashed lines represent the T-FT setting, where only the task model is fine-tuned.}\label{fig:baselineCOCOold}
\end{figure}

\begin{figure}[t]
	\centering
	\includegraphics[trim=10 10 20 40,clip,width=0.7\textwidth]{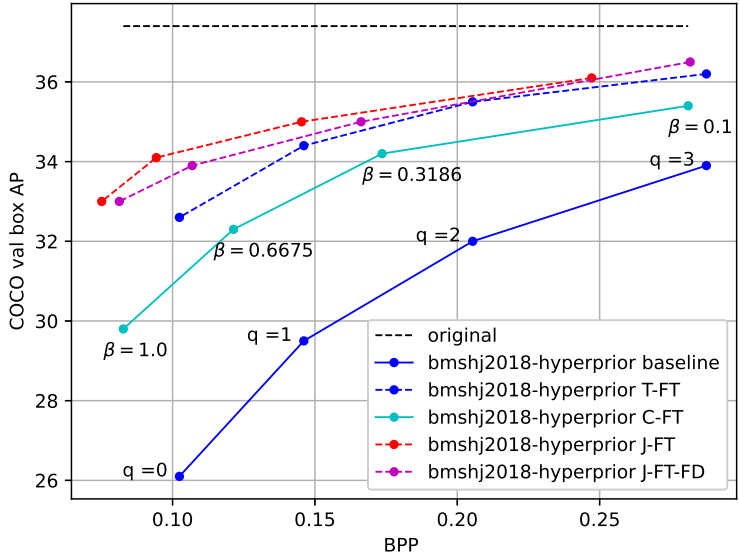}
	\caption{Rate-accuracy curves for object detection. The top dashed line represents the original detection accuracy on reference images ($rq$).}\label{fig:cocov4_DF}
\end{figure}

\begin{figure}[t]
	\footnotesize
	\setlength{\tabcolsep}{1pt}
	\centering
	\begin{tabular}{llll}
		Original & T-FT&C-FT &J-FT\\	
		\includegraphics[width=0.24\linewidth]{figures/sameAP/original.jpg.eps}&
		\includegraphics[width=0.24\linewidth]{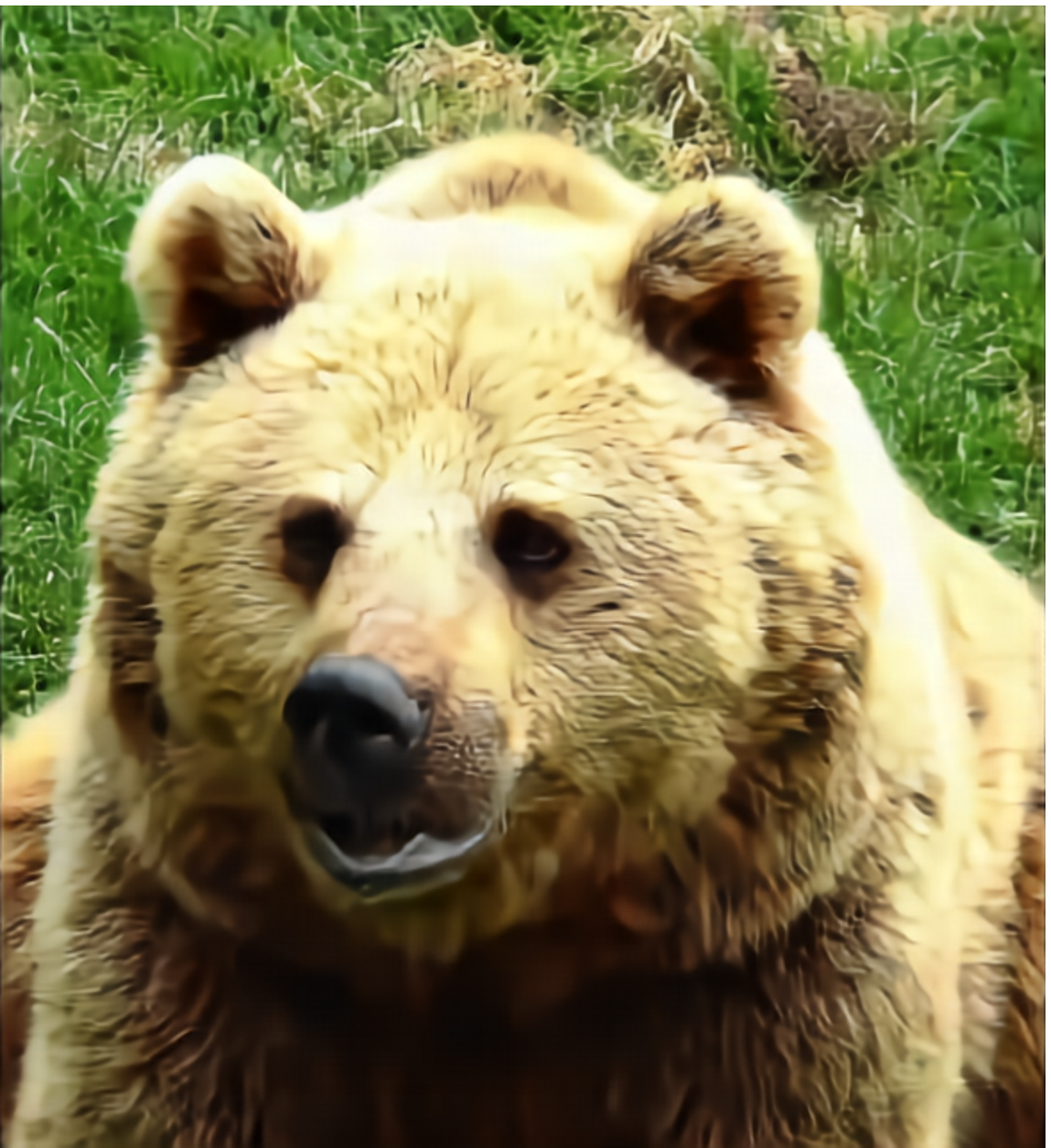}&
		\includegraphics[width=0.24\linewidth]{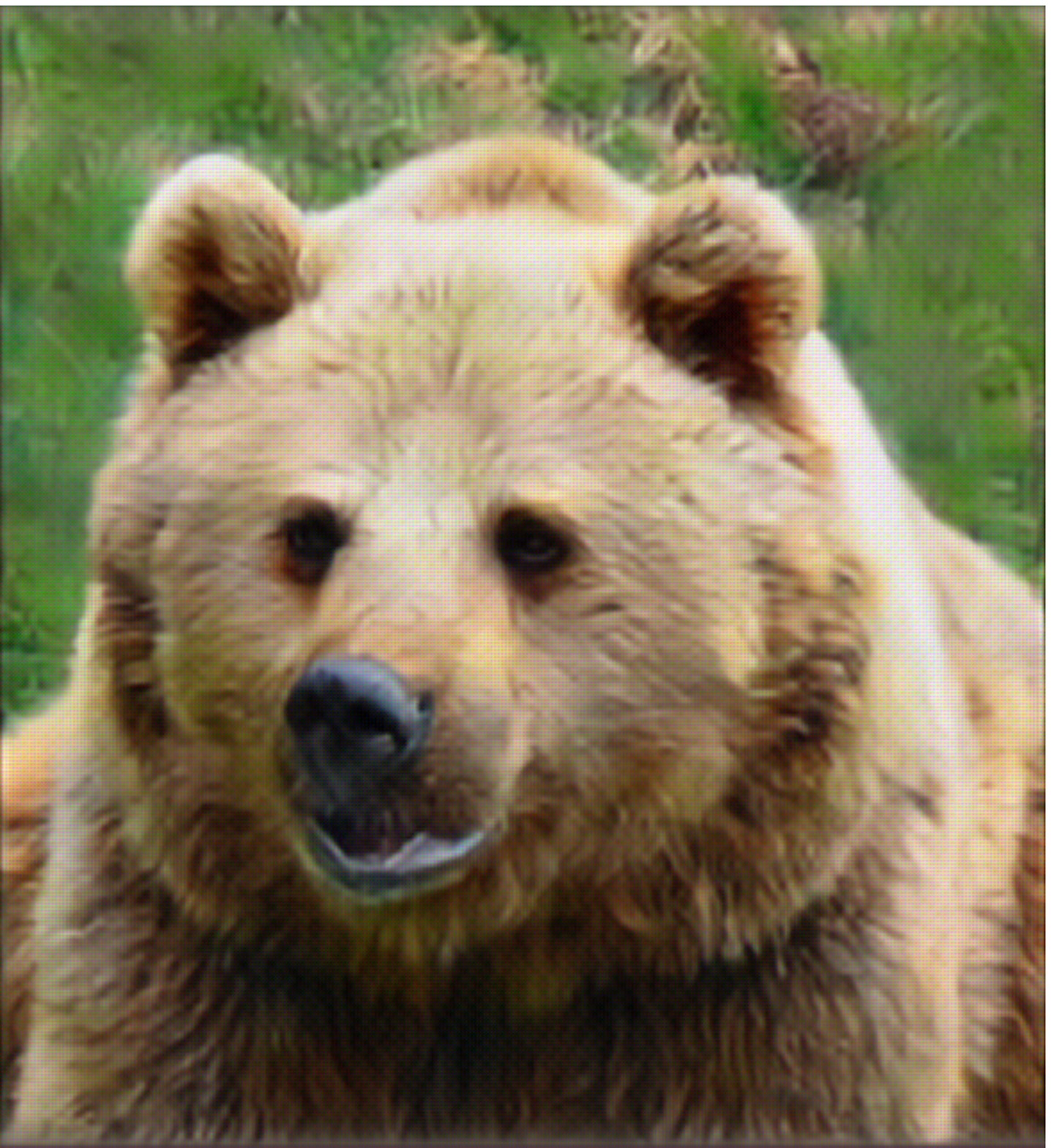}&
		\includegraphics[width=0.24\linewidth]{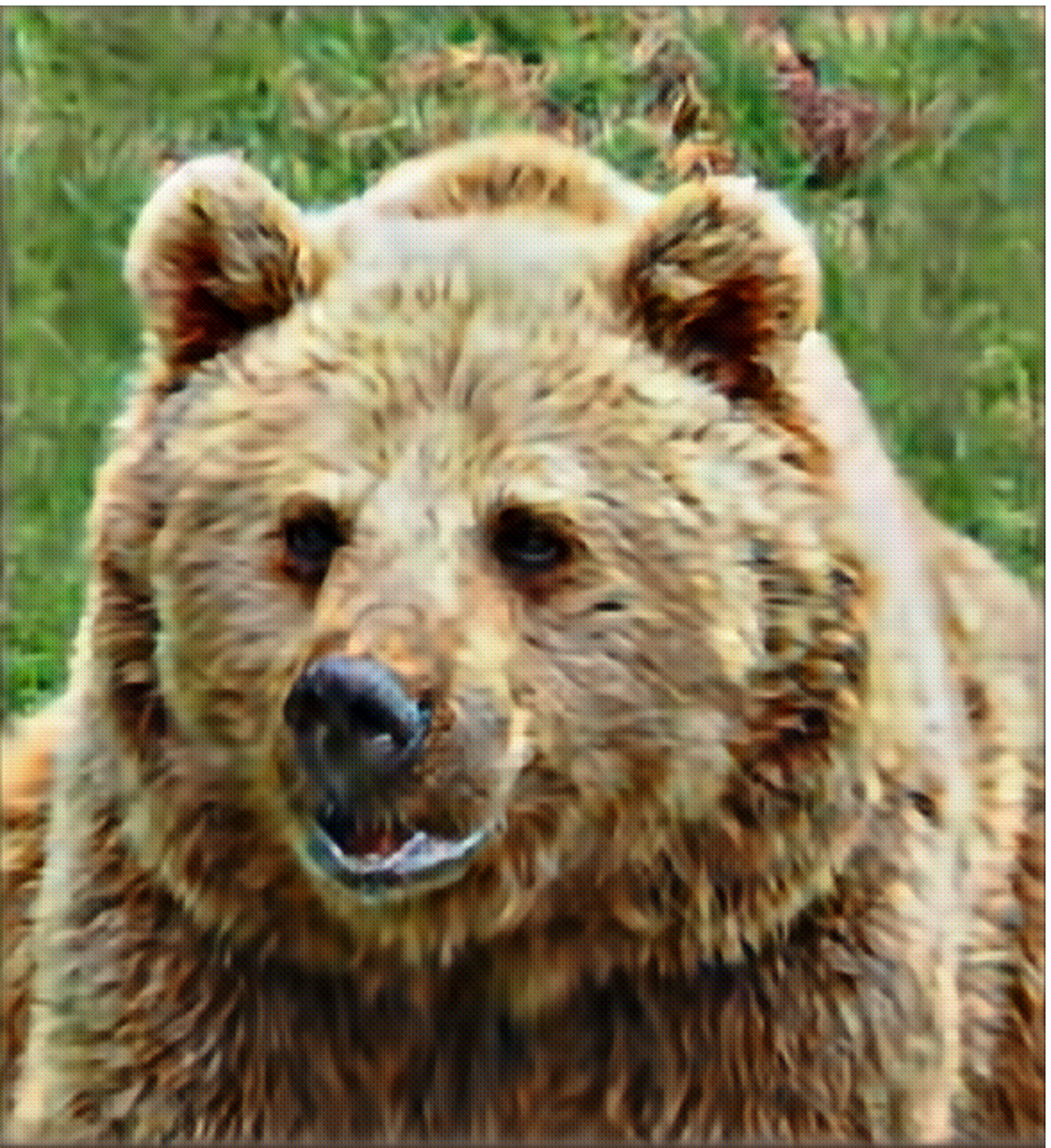}\\
		box mAP        & 32.6   & 29.8   & 34.1 (\textbf{+4.6\%)} \\
		PSNR (dB)      & 30.49  & 17.70  & 16.31                 \\
		bpp            & 0.1024 & 0.0827 & 0.0943
	\end{tabular}
	\caption{Reconstructed `bear' at constant bit-rate $\approx$ 0.1bpp. J-FT improves the detection accuracy by 4.6\% compared to the task-only training setting.}\label{fig:reconCompare2}
\end{figure}

\begin{figure}[t]
	\footnotesize
	\setlength{\tabcolsep}{1pt}
	\centering
	\begin{tabular}{llll}
		Original & T-FT&C-FT &J-FT\\	
		\includegraphics[width=0.24\linewidth]{figures/sameAP/original.jpg.eps}&
		\includegraphics[width=0.24\linewidth]{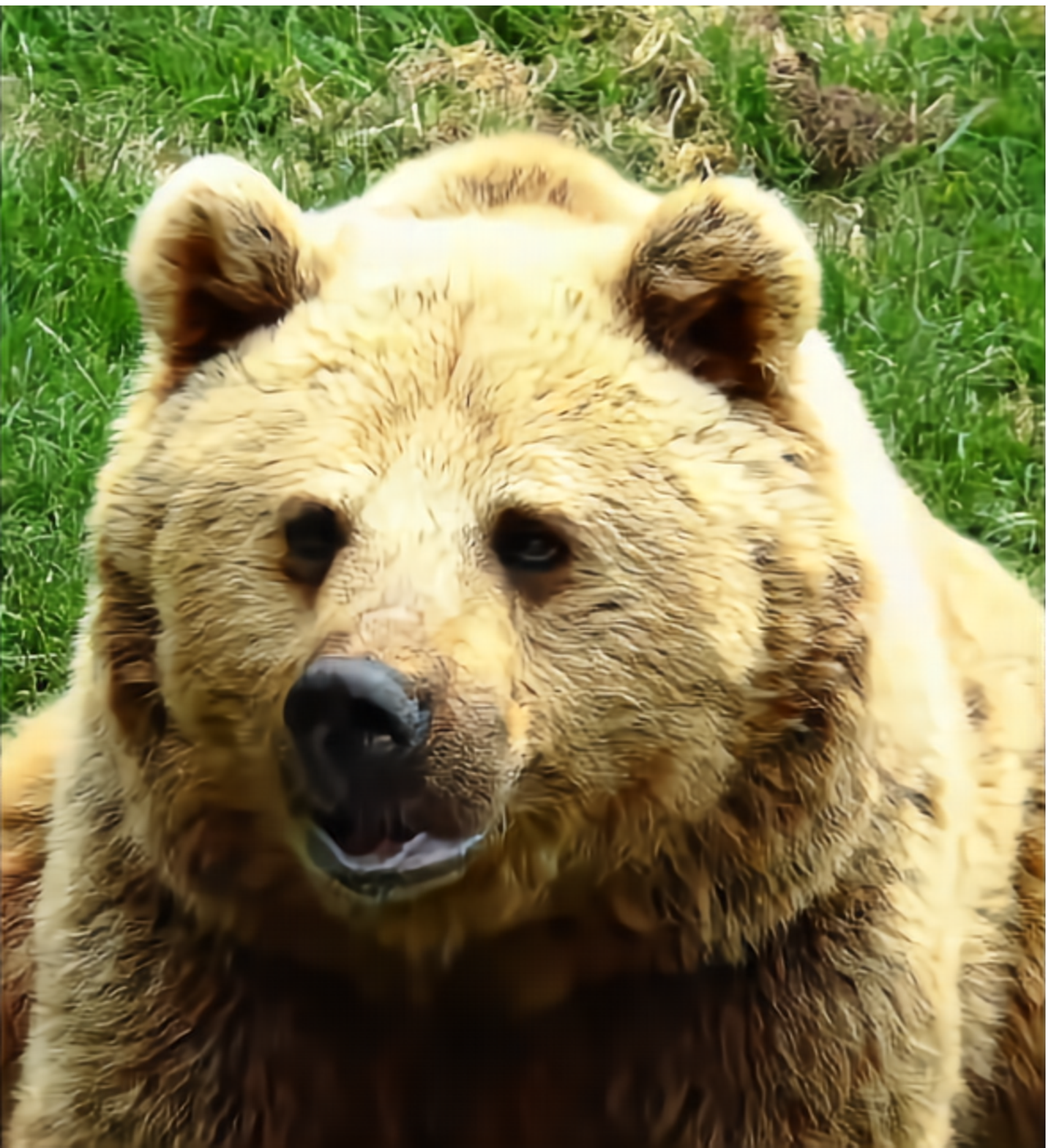}&
		\includegraphics[width=0.24\linewidth]{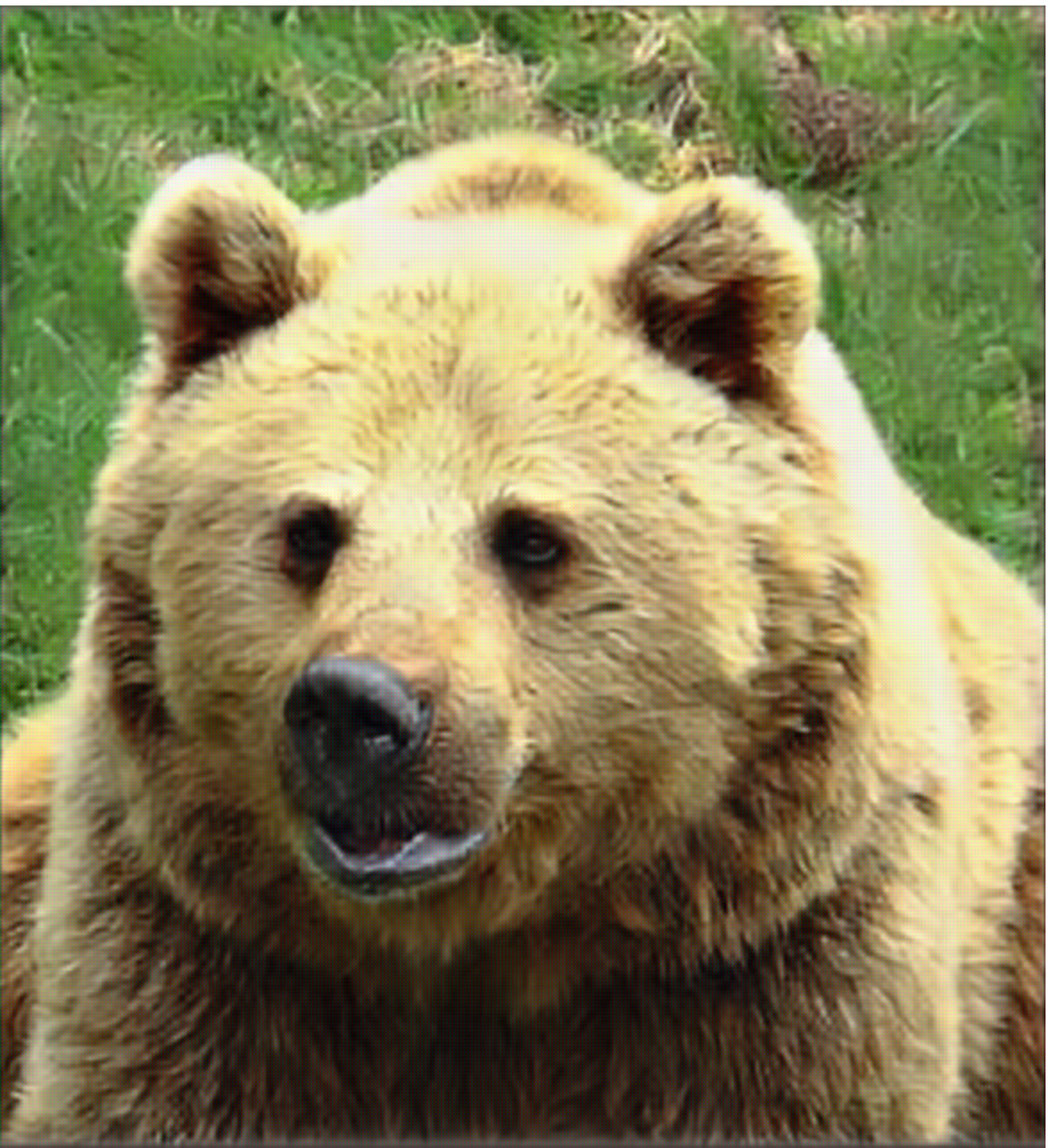}&
		\includegraphics[width=0.24\linewidth]{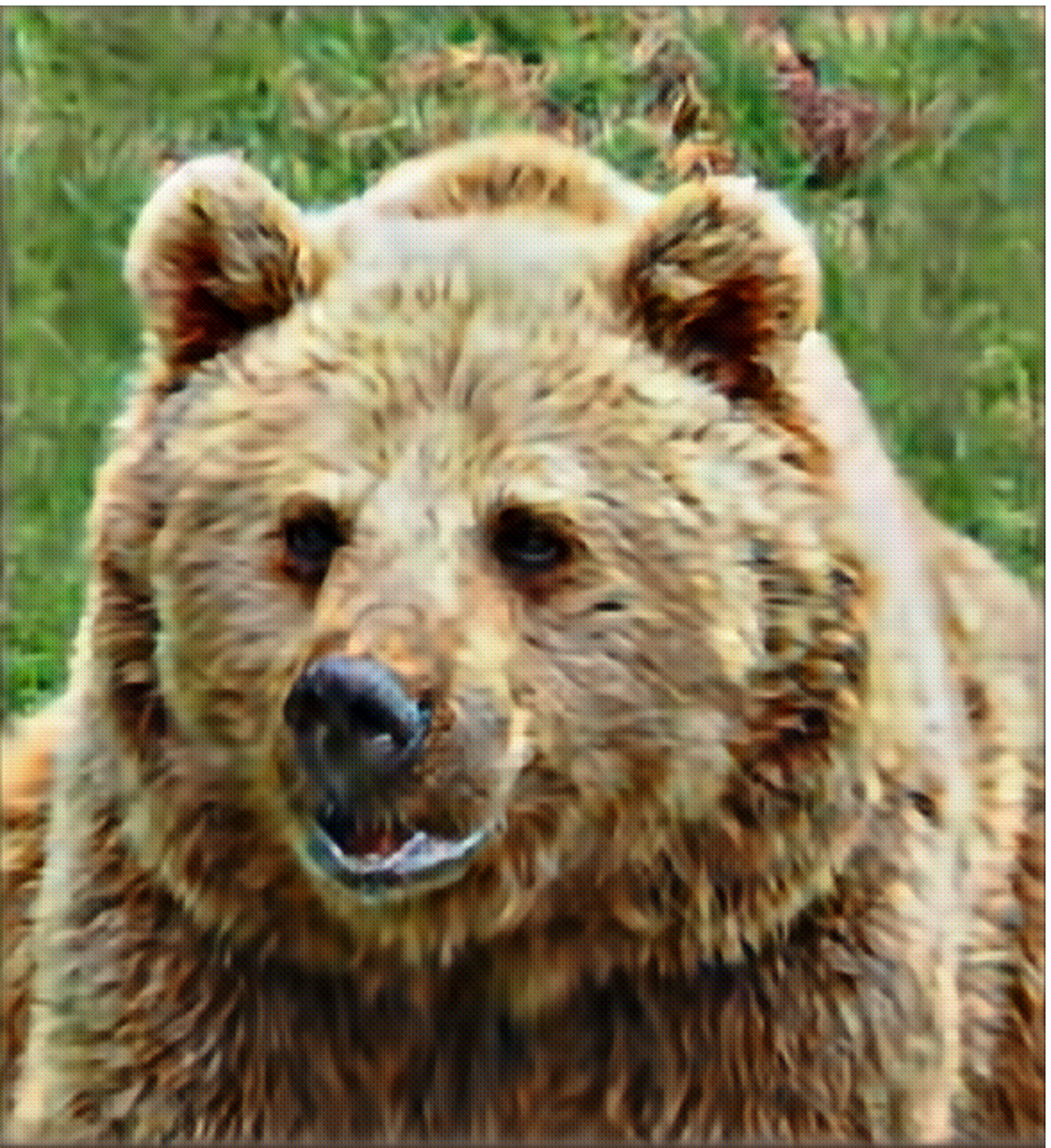}\\
		box mAP   & 34.4                           & 34.2                           & 34.1   \\
		PSNR (dB) & 32.11                          & 20.01                          & 16.31  \\
		bpp       & 0.1461 (\textbf{+54.9\%})      & 0.1736 (\textbf{+84.1\%})      & 0.0943 \\
	\end{tabular}
	\caption{Reconstructed `bear' at similar detection accuracy box mAP $\approx$ 34-34.5.}\label{fig:reconCompare1}
\end{figure}

\begin{figure}[t]
	\centering
	\subfloat[]{\includegraphics[trim=10 10 40 20,clip,width=0.50\textwidth]{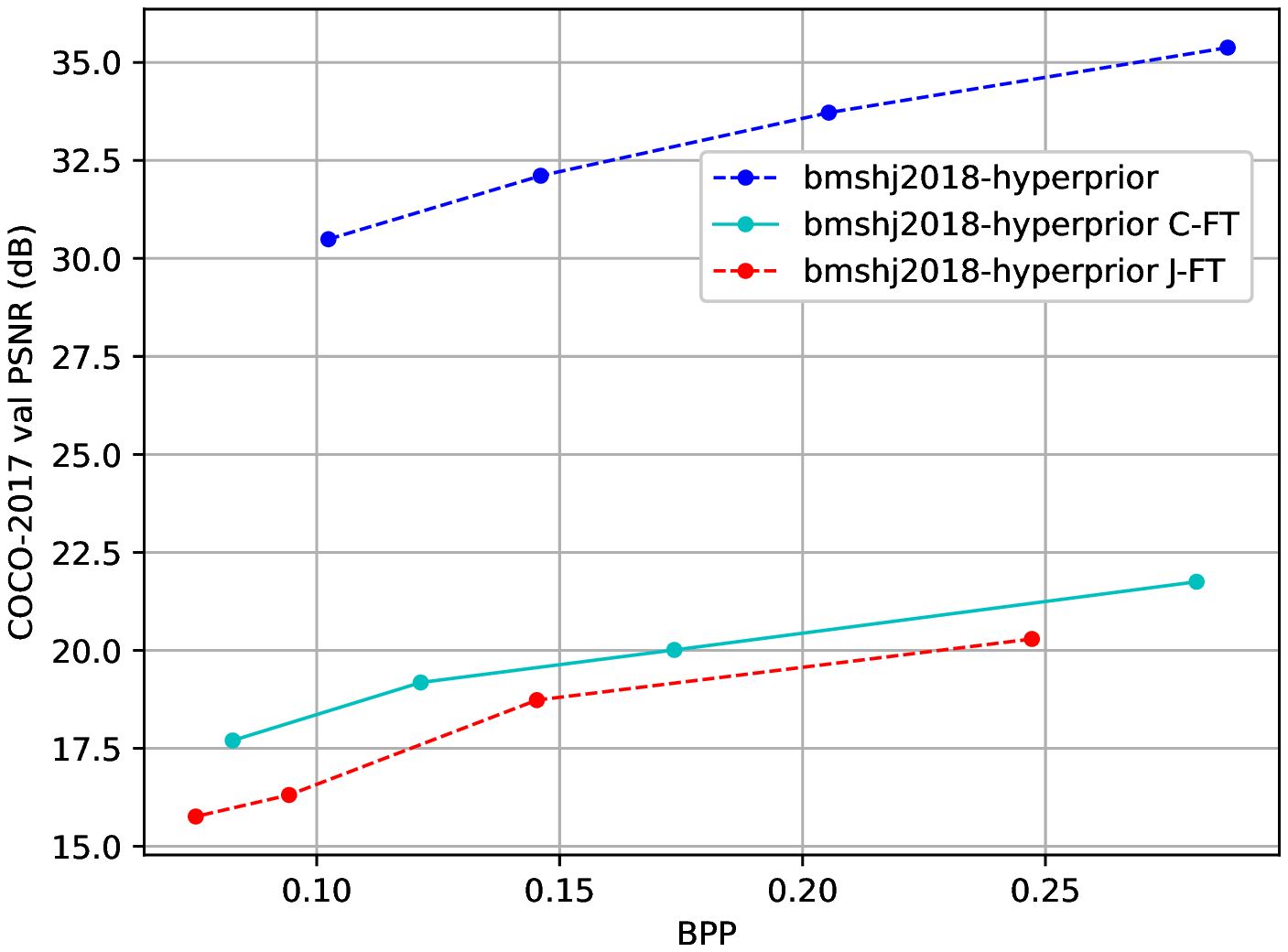}}\hfill
	\subfloat[]{\includegraphics[trim=10 10 40 20,clip,width=0.50\textwidth]{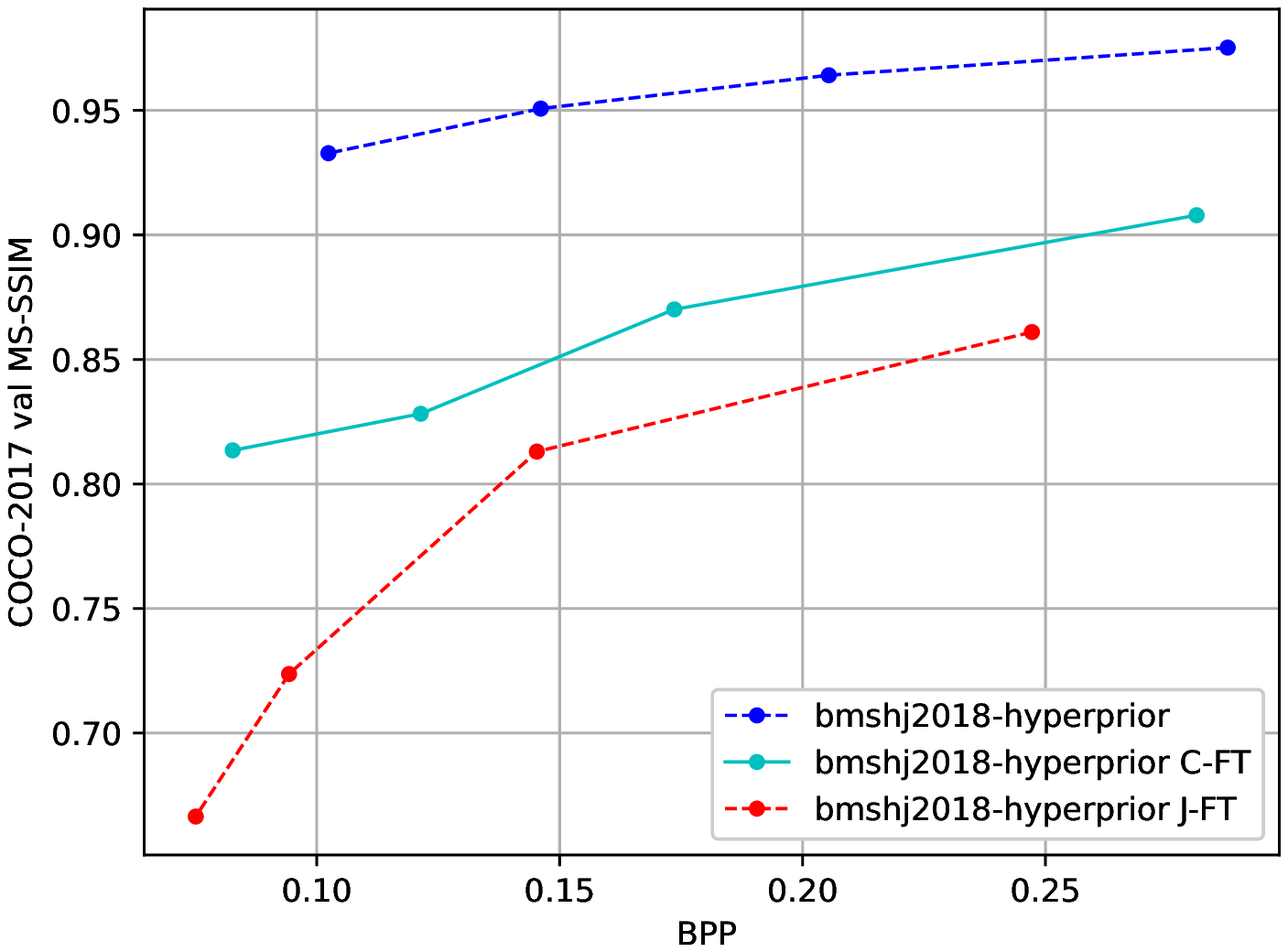}}
	\caption{PSNR (a) and MS-SSIM (b) rate-distortion curves for the studied configurations, measured on the COCO-2017 validation set.}\label{fig:ratedistortion}
\end{figure}

\subsection{Experimental setup}\label{sec:exp:setup}

To test the previously described optimization schemes, we use the well researched object detection task, referred as $T$, bench-marked on the COCO-2017 dataset~\cite{lin2014microsoft}. This dataset contains over 180k training and 5k validation RGB images of average size over 256$\times$256 with annotated bounding boxes belonging to 80 classes.
To evaluate the performances under different settings and bit-rate constraints we use the primary challenge metric\footnote{\url{https://cocodataset.org/\#detection-eval}}, denoted `box mAP', to evaluate the detection accuracy.
As for the detection model, we use the Faster-RCNN~\cite{ren2015faster} model with a ResNet-50 backbone throughout the experiments.

H.266/VVC and the scale hyperprior model \textit{bmshj2018-hyperprior}~\cite{balle2018variational}, optimized for Mean Square Errors (MSE), are used as baseline codecs. H.266/VVC is the latest state-of-art image/video compression standard from MPEG/ITU. For VVC, we used the reference software VTM version 8.2\footnote{Available at \url{https://vcgit.hhi.fraunhofer.de/jvet/VVCSoftware_VTM}} in still picture mode to compress the images (\textit{all intra} configuration). For \textit{bmshj2018-hyperprior}, we use the hyperprior configuration described by Ball\'{e} \textit{et.~al} in~\cite{balle2018variational}, with the implementation and pre-trained weights provided by CompressAI~\cite{begaint2020compressai}. The actual bit-rates are reported, measured on the final bit-stream encoded by the entropy coder.

The rate-accuracy curves of VVC are generated using the following quantization parameters (QP): 47, 42, 37, 32, 27, 22, 17 (the quality increases with decreasing QP). The rate is represented in bits per pixel (bpp), derived from the total size of the bit-stream and the spatial dimensions of the images. We use the quality parameter $q\in [1,8]$ to load different pre-trained weights from~\cite{begaint2020compressai} and generate images at different rates. 

\subsection{Task model fine-tuning (T-FT)}\label{sec:exp:tft}
To observe the performance gain for the T-FT setting as given in Eq.~\eqref{eq:tft}, we fine-tuned the Faster-RCNN model starting from the pre-trained parameters with a stochastic gradient optimizer at a learning rate of $0.01$ for 6 epochs.
For each rate-accuracy point for both VVC and \textit{bmshj2018-hyperprior}, we fine-tuned
the task model with the respective decompressed images. In Figure~\ref{fig:baselineCOCOold}, we compare the performance gains for VVC and \textit{bmshj2018-hyperprior} with and without fine tuning the task model. Fine-tuning the task algorithm on reconstructed images unsurprisingly improves rate-accuracy trade-offs, especially at low bit-rates. Interestingly, in both cases the two codecs perform similarly in terms of rate-accuracy, while the \textit{bmshj2018-hyperprior} cannot compete against VVC in PSNR~\cite{begaint2020compressai}.

\subsection{Codec fine-tuning (C-FT)}\label{sec:exp:cft}

In this configuration, the \textit{bmshj2018-hyperprior} model is fine-tuned alongside a fixed pre-trained detector to minimize the rate-accuracy loss given in Eq.~\eqref{eq:cft}. For each desired rate-accuracy point, we fine-tuned the codec from pre-trained weights from~\cite{begaint2020compressai} (originally optimized for MSE) at each quality $q$ and selected a control parameter $\beta$ accordingly, which denotes the Lagrangian multiplier in the rate-accuracy loss function (Eq.~\ref{eq:cft}). 
Table~\ref{tab:cft} lists the corresponding hyper-parameters $q$ and $\beta$ used for initialization and training. The codec is fine-tuned for 6 epochs on the COCO-2017 training set, using the Adam optimizer, a batch size of 2 and a learning rate of $10^{-4}$.

Figure~\ref{fig:cocov4_DF} compares the results of the proposed settings. One can notice that retraining the codec only (C-FT) underperforms task fine-tuning (T-FT), but still improves significantly over the off-the-shelf codec.

\begin{table}[t]
	\caption{\label{tab:cft}Hyper-parameters used for fine-tuning the codec in C-FT and J-FT configurations.}
	\centering
	\begin{tabular}{lcccc}
		\toprule
		Codec quality ($q$) & 1 & 2 & 4 & 4\\
		\midrule
		Control parameter ($\beta$) &1.0&0.6675&0.3186&0.1\\
		\bottomrule
	\end{tabular}
\end{table}

\subsection{Joint end-to-end fine-tuning (J-FT)}

In this setting, the compression model and the detector are jointly optimized by solving Eq~\eqref{eq:jft}. We fine-tuned the compression model and the detector for 5 epochs with the same optimizer and learning rates used in Sections~\ref{sec:exp:cft} and~\ref{sec:exp:tft}. Since \textit{bmshj2018-hyperprior} contains less parameters to train than the Faster-RCC model and converges faster, we fine-tuned the detector for one additional epoch (6 epochs total). We observed that this strategy gains over 7.1\% and 1.6\% detection accuracy at 0.08bpp and 0.25bpp, respectively.
It can be seen in Figure~\ref{fig:cocov4_DF} that Joint Fine-Tuning (J-FT) logically outperforms the other training configurations. For example, we observe over 4.5\% improvement compared to T-FT at 0.1bpp.

In order to get an idea of the performance of a fixed decoder, in the context of standardizing a bit-stream dedicated to image/video coding for object detection, we performed one final experiment where we fine-tuned joint task algorithm and auto-encoder, without updating the decoder parameters. This is referred to as J-FT-FD for Fixed Decoder in Figure~\ref{fig:cocov4_DF}, which shows interesting properties since the performance does not drop significantly compared to J-FT. 

\subsection{Visual quality of decompressed images}

In Figure~\ref{fig:reconCompare1} and Figure~\ref{fig:reconCompare2}, we compare the visual quality of the reconstructed image `bear' from the COCO-2017 validation set, at similar bit-rates and similar detection accuracy, respectively. Since the loss functions in Eq.~\eqref{eq:cft} and \eqref{eq:jft} do not contain any distortion terms on the reconstructed images, settings C-FT and J-FT are clearly not optimized for viewing. One can notice that the reconstructed images for the settings J-FT and C-FT highlight the features which help improving the object detection accuracy. To analyze the amount of distortion introduced by the proposed settings quantitatively, Figure~\ref{fig:ratedistortion} shows rate-distortion curves, measured in PSNR and MS-SSIM, on the COCO-2017 validation set, of \textit{bmshj2018-hyperprior} in C-FT and J-FT configurations, compared with the pre-trained models optimized for MSE. The re-training of the codec clearly impacts decoded pictures' fidelity to the source, although the images are still viewable for visual control.
\section{Conclusion}\label{sec:conclusion}

This paper detailed an end-to-end framework enabling efficient image compression
for remote machine task analysis, using a chain composed of a compression module
and a task algorithm that can be optimized end-to-end. The experiments demonstrate that it is possible to significantly improve the task accuracy when fine-tuning jointly the codec and the task networks, especially at low bit-rates. Depending on training or deployment constraints, selective fine-tuning can be applied only on the encoder, decoder or task network and still achieve rate-accuracy improvements over an off-the-shelf codec and task network. This also demonstrates the flexibility of end-to-end pipelines for practical applications.

ANN-based codecs have already been proven to reach performances close to conventional methods, and to outperform them when trained with perceptual quality metrics~\cite{minnen2020channel,balle2018variational,minnen2018joint,mentzer_high-fidelity_2020}. This paper highlights another clear advantage of ANN-based codecs: optimized compression for machines. In future work, we plan to study if the promises of task-based optimized compression will hold for video applications and extend ANN-based codecs to support multiple types of machine tasks.

\newpage
\Section{References}
\bibliographystyle{IEEEbib}
\bibliography{ms}

\end{document}